# Solving combinatorial problems by two D-Wave hybrid solvers: a case study of traveling salesman problems in the TSP Library

Richard H. Warren

Lockheed Martin Corporation (Retired), King of Prussia, PA 19406, USA

rhw3@psu.edu

**Abstract**

The D-Wave quantum computer is an analog device that approximates optimal solutions to optimization problems. The traveling salesman problems in the TSP Library are too large to process on the D-Wave quantum computer DW_2000Q_6. We report favorable approximations for solving the smallest, symmetric traveling salesman problems in the TSP Library by two D-Wave hybrid solvers, Kerberos and LeapHybridSampler. This is useful work about results from new quantum tools on problems that have been studied. It is expected to show a quantum way forward with larger problems when the hardware is upgraded. Also this work demonstrates that the TSP Library is a source of benchmarks for quantum processing of combinatorial problems.

The hybrid solvers combine quantum and classical methods in a manner that is D-Wave proprietary information. The results from Kerberos were closer to optimal than the results from LeapHybridSampler. We show that the error percent from optimal increases as the problem size increases, which is consistent with results on D-Wave's quantum computer for other optimization problems.

An appendix contains outcomes from the two hybrid solvers for two asymmetric traveling salesman problems that are in the TSP Library. Again, the Kerberos results were closer to optimal than those from LeapHybridSampler, which indicates that Kerberos is superior to LeapHybridSampler on traveling salesman problems.

## 1. Introduction

We studied Burma14, Ulysses16, gr17 and Ulysses22 which are traveling salesman problems (TSPs) from the TSP Library [1] with 14, 16, 17 and 22 cities, respectively. We chose them because they are symmetric (asymmetric TSPs are more difficult to solve, in general), have been studied [2, 3], are the smallest TSPs in the library, and are too large to process on the D-Wave quantum machine DW_2000Q_6 that has 2,030 qubits, but six or less connections per qubit. We report about the ability of D-Wave's hybrid solvers Kerberos [4] and LeapHybridSampler (LHS) [5] to find near-optimal tours for these TSPs. The solvers are hybrid in the sense that they employ quantum and classical techniques. The hybrid methods are not available since they are D-Wave trade secrets.

Outstanding results for solving four asymmetric 10-city traveling salesman problems by the D-Wave hybrid solver Kerberos are reported in [6]. The test made four executions for each problem. All Kerberos solutions were optimal tours.

The results that we are reporting in this paper are less favorable, but expected because as optimization problems grow in size, they become more difficult to solve on a quantum computer [2]. The Kerberos results are closer to the length of an optimal tour than the results from LHS. This implies that Kerberos is preferred to LHS on traveling salesman problems.

This paper begins with preliminaries in Section 2 about TSPs, quantum annealing, solvers and error percent. Section 3 is about processing TSPs with a hybrid solver. In Section 4 each of the four TSPs that we study has a subsection about its solver results. Section 5 has global results and conclusions. Section 6 contains an outlook into the future. An Appendix contains solver outcomes for the two smallest, asymmetric TSPs that are in the TSP Library.

## 2. Preliminaries

The TSP is famous because it is easy to express, has NP-complete complexity, and has numerous applications. Given N cities and the distance between city i and city j, the symmetric TSP [7] asks for a shortest route through the N cities visiting each city once and returning to the originating city. When the distance from city i to city j may be different than the distance from city j to city i, it is called an asymmetric TSP (ATSP). The adaption of the TSP to a quantum annealer [8, 9] is for the ATSP. This is natural and prevents subloops through proper subsets of the N cities. A tour for the salesman is one loop through all the cities, i.e., a cyclic permutation of the cities. The length of a tour is the distance traveled through the loop. An optimal tour is a shortest tour.

Quantum computing is based on the properties of quantum mechanics. It has received major attention since the mid-nineteen-nineties when Peter Shor proved theoretically that integer factorization on a quantum computer can be accomplished faster than on a classical computer. In general, quantum computing has two major branches; the discrete technique called gate model and the continuous technique often called annealing. It has been shown that the two techniques are equivalent, i.e., any algorithm that can be implemented in one of the techniques can be translated to the other technique.

Quantum annealing [10] is a method of computing that has the potential to solve optimization problems faster than classical means. The theory for quantum annealing implies that the qubits will achieve an optimal state of low energy when super cooled. This is represented by Eq. (1) where an initial Hamiltonian $H_o$ evolves to its low energy state in a final Hamiltonian $H_p$ according to

$$H(t) = \left(1 - s\left(\frac{t}{T}\right)\right) H_o + s\left(\frac{t}{T}\right) H_p \text{ for } 0 \leq t \leq T \tag{1}$$

as $s(t/T)$ increases from $s(0) = 0$ to $s(1) = 1$ and if $H_o$ and $H_p$ do not commute. In theory, $T$ is the time imposed by the Schrödinger equation for the initial Hamiltonian to evolve to its low energy state. On a D-Wave computer, time $T$ is in microseconds. The Hamiltonian $H_o$ is established by D-Wave for all problems. The Hamiltonian $H_p$ represents the combinatorial problem to be solved and is an input. Besides $H_p$, other inputs include the number of times the problem is solved, the time $T$ within a given range, and scaling factors. Essentially, a result $H(t)$ is a sample from a Boltzmann distribution. A physical implementation of Eq. (1) does not strictly meet the conditions for quantum annealing. Noise disrupts the quality of calculations. In addition, values loaded by the user may differ slightly from the machine interpretation of the numbers. These difficulties are partially overcome by taking multiple samples and choosing a valid solution that has minimum energy.

There is a significant complication solving TSPs on the D-Wave quantum processor DW_2000Q_6. The number of connections needed between logical qubits, which is O((N – 1)! x (N – 2)) for a symmetric TSP on N cities, may exceed the number of available connections which is at most 6 per physical qubit. My experience on the DW_2000Q_6 processor is that symmetric TSPs transition between solvable/unsolvable for about N = 8. This difficulty is circumvented by using a hybrid solver which breaks the TSP into manageable pieces, solves them, and combines the solutions to obtain the best overall solution.

Kerberos [4] is a D-Wave hybrid problem solver. D-Wave claims it can unravel a problem of arbitrary structure and size asynchronously by classical tabu search and simulated annealing in parallel. In each iteration, best results are sent to the quantum processing unit for sampling. The lowest energy result is returned as the best solution for that iteration.

LeapHybridSampler (LHS) [5] is a more recent D-Wave hybrid problem solver. D-Wave claims it can handle problems with up to 10,000 variables automatically by running them on quantum and classical resources. D-Wave's algorithms which are trade secrets decide the best way to solve a problem.

For each combination of TSP and solver, we measure closeness to optimal by

Error Percent = [(Best Solution – Optimal Solution) / Optimal Solution] x 100     (2)

where Error Percent is the percent of relative distance from the optimal tour. Best Solution is the shortest tour length achieved by experimentation. Optimal Solution is the length of an optimal tour.

## 3. Method to solve TSPs with a D-Wave hybrid solver on a quantum computer

We outline our design for solving TSPs by a D-Wave hybrid solver on quantum processor DW_2000Q_6. Comments and explanations are included.

1. Set problem size N = number of cities.

2. Import N x N distance matrix {$d_{ij}$} where $d_{ij}$ is the distance from city i to city j. The diagonal entries have no role.

3. Designate the N² variables which are city *k* in position *m* of a tour. Due to redundancy, we may assume that city 1 is always in position 1 which reduces the number of variables to (N – 1)². A helper function is needed to reduce the double subscripts to one index.

4. Set tunable parameters

Gamma is the Lagrange parameter. It balances constraints and the objective. Constraints ensure each city on the route is visited exactly once. The objective is to find a route with shortest distance traveled.

Chainstrength ensures the embedding function binds together physical qubits to represent one logical qubit. If it is overly strong, it may bind other qubits.

Numrums is the number of times the hybrid algorithm is to run the problem. Its range is 10 – 1000. Since the D-Wave processor is analogue, it needs to be large enough to run the problem many times looking for the best solution, but not excessively large to use too much processing time.

Time *T* in Eq. (1) within a given range for annealing. It needs to be high enough for the ground state to progress to its final state, but not unduly high so that too much processing time is used.

5. Set the objective function which minimizes the distance traveled. It is equation (2) in [8] and repeated as equation (56) in [9].

6. Set the constraint functions that ensure each city on the route is visited exactly once. They are equations (3) and (4) in [8] and combined as equation (57) in [9]. The sum of the objective function and the constraint functions is the Hamiltonian $H_p$ in Section 2.

7. Import embedding function that matches variables in the problem with qubits.

8. Import hybrid application that breaks problem into parts that can be solved in a classical-quantum arrangement and selects best overall solution from the pieces.

9. The output is a list of selected qubits and their position in the route. External verification is needed to ensure that the list represents a valid tour. The length of the tour needs to be determined.

## 4. Results

We present the data outcomes for the four smallest, symmetric TSPs in the TSPLIB. All of them are too large to process on the D-Wave quantum device DW_2000Q_6 due to an insufficient number of connections between qubits.

### 4.1. Burma14

Although Burma14 is a symmetric TSP with 14 cities, it seems difficult to solve for an optimal tour by D-Wave solvers. Some of the obstacles include a large difference of 1,242 between the largest and smallest distances between two cities, a small gap of 13 between the optimal tour length of 3323 [1] and the next smallest tour length, and only two optimal tours (1 2 14 3 4 5 6 12 7 13 8 11 9 10 1) and its inverse. In general, prior quantum work on D-Wave [11] has not been able to distinguish between solutions with small variance, and when there is a unique optimal answer.

The D-Wave solver Kerberos did not produce an optimal tour for Burma14. The best solution using Kerberos had tour length 3545. Error Percent, defined by Eq. (2), is about 6.7% from the optimal tour. We executed Kerberos eight times on Burma14, seven with the default settings which include number of iterations equal to 100, one with this setting at 1000. The change seemed to have no effect.

The D-Wave solver LHS was worse than Kerberos on Burma14. In fact, each result from LHS is further from the optimal tour length than the results from Kerberos.

The distance matrix for Burma14 is in [12]. The Concorde TSP solver [13] found an optimal tour for Burma14 in .02 seconds, which is not surprising since Concorde is designed to solve symmetric TSPs extremely fast.

## 4.2. Ulysses16

To ensure that our software was valid for 16-city TSPs, we tried a simple, symmetric 16-city TSP for which we knew the optimal tour. The lowest energy samples were not optimal, but we showed that the software would produce a tour and place city 1 in position 1 of a tour.

Ulysses16 is symmetrical with 16 cities. The maximum and minimum distances between two cities are 2789 and 52. An optimal tour has length 6859 [1]. Apparently, the closest tour to an optimal tour has length 6865. An optimal tour is (1 8 4 2 3 16 10 9 11 5 15 6 7 12 13 14 1).

The D-Wave solver Kerberos did not find an optimal tour for Ulysses16. The best solution by Kerberos had tour length 7314 which is an Error Percent of 6.6% from an optimal tour. We executed Kerberos six times on Ulysses16, all with the default settings.

The D-Wave solver LHS was not as good as Kerberos on Ulysses16. Five of the six results from LHS were further from the optimal tour length than the results from Kerberos. The sixth result was slightly closer to optimal than the worst result from Kerberos.

## 4.3. gr17

The symmetric TPS gr17 with 17 cities does not appear to be fully studied, since we could not find an optimal tour in the literature. It has largest distance 745 and smallest distance 27. The length of an optimal tour is 2085 according to [1]. The distance matrix is in [14].

We successfully executed Kerberos six times. None of the results were optimal. Error Percent, defined by Eq. (2), was 13.86%.

Also, we successfully executed LHS six times. All results were longer than the Kerberos results. The Error Percent for LHS on gr17 was 53.53%.

### 4.4. Ulysses22

Ulysses16 a subset of Ulysses22. The first 16 cities in Ulysses22 are identical to those in Ulysses16 and are listed in the same order.

The longest distance in Ulysses22 is 2789 and the shortest is 14. An optimal tour has length 7013 according to [1]. A distance matrix is found in [15].

We successfully executed the solver Kerberos six times on Ulysses22. None of the results were optimal. The Error Percent was 16.50%.

We performed twelve executions using the solver LHS. The output data for eight executions was corrupt and is not reported here. We attempted to correct this for three executions by increasing the minimum time for execution from 3 seconds to 10 seconds, then to 50 seconds, and finally to 300 seconds. It seemed to have no effect. This is the only LHS parameter available to a user. The TSP portion of the algorithm has one parameter; the tunable parameter gamma for controlling constraints in the quantum formation of the TSP. Initially it was set to N x max($D_{xy}$) / 2 = 22 x 2,789 / 2 = 30,679. When reduced, it allowed four solutions with slightly corrupted data. These four solutions have longer lengths than the Kerberos solutions.

### 5. Discussion and Conclusions

Fig. 1 shows the optimal tour length and the tour lengths obtained by Kerberos and LeapHybridSampler for the four TSPs that we studied. Fig. 2 shows the Error Percent for Kerberos and LeapHybridSampler for the four TSPs. Table 1 contains solution data for the four TSPs.

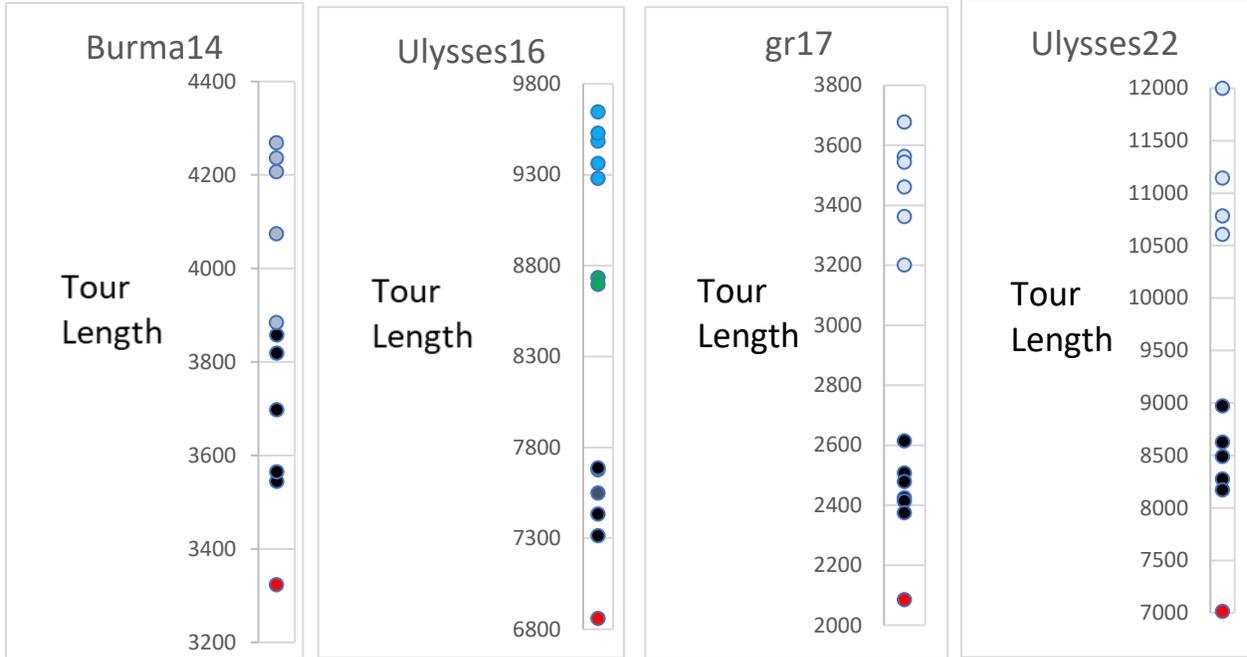

Fig. 1. (Color online) Red dots are optimal tours. Black dots are Kerberos results. Blue dots are results from LeapHybridSampler.

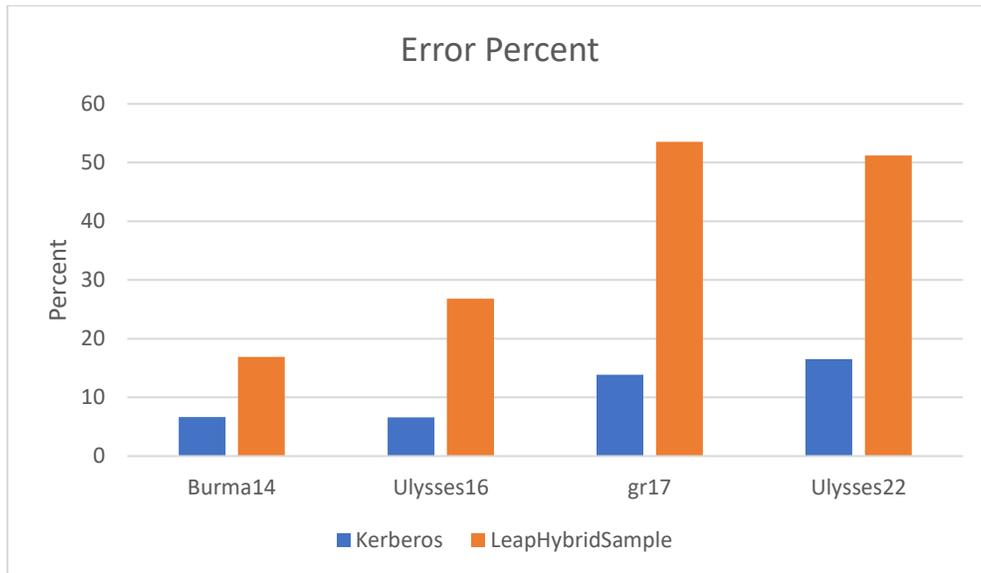

Fig. 2. (Color online) Error Percent is defined by Eq. (2). Smaller percent is better because it is closer to optimal.

Table 1. Results for the Traveling Salesman Problems

| TSP | Optimal Solution | Best Kerberos Solution | Error Percent % | Best LeapHybrid Sampler Solution | Error Percent % |
|---|---|---|---|---|---|
| Burma14 | 3323 | 3545 | 6.7 | 3884 | 16.9 |
| Ulysses16 | 6859 | 7314 | 6.6 | 8697 | 26.8 |
| gr17 | 2085 | 2374 | 13.9 | 3201 | 53.5 |
| Ulysses22 | 7013 | 8170 | 16.5 | 10605 | 51.2 |

Optimal solutions are from [1]. Error Percent is determined according to Equation (2).

We conclude from Figures 1 and 2 and from Table 1 that Kerberos is a superior solver to LeapHybridSampler on the four TSPs studied in this paper. Also, as the number of cities increases, the Error Percent, defined by Eq. (2), tends to increase for both solvers on the four TSPs. This agrees with an observation in [2] for a different solver.

This work supports the conclusions in [16] that the symmetric TSPs in the TSP Library [1] are excellent candidates to serve as benchmarks for combinatorial optimization problems. This is evident, not only by our work, but by the availability of the TSP Library and the cited results in the literature about these symmetric TSPs.

## 6. Future Perspectives

The quantum challenge for the TSP is to solve large problems rapidly and close to optimal. This is a distant goal that can be achieved mainly by hardware improvements.

D-Wave announced a hardware upgrade [17] that will increase the number of qubits from 2,000 to 5,000 and have more connectivity between qubits. The expectation is that symmetric TSPs with more than 8 cities can be processed by the new hardware and analyzed quantumly without a hybrid solver. Then hybrid solvers can be tried on the larger TSPs investigated by Feld et al. and reported in [2]. They are WesternSahara29 and Djbout38. Other TSPs from the TSPLIB, including asymmetric ones, can be included in future hybrid studies.

**Funding:** The author declares that no funding was provided for this effort.

**Data Availability Statement:** The data and scripts that support the findings of this study are available from the author upon reasonable request.

**Acknowledgements:** David Johnson of D-Wave Systems is thanked for technical guidance about Kerberos and LeapHybridSampler. Patrick Fleury is thanked for help with Python in the scripts. Jesus Christ is thanked for being my personal savior and promising me a future place with Him.

**Conflicts of Interest**: The author declares no conflict of interest.

## Appendix:  Examining Two Asymmetric Traveling Salesman Problems in the TSP Library

### A1.  Introduction

We continue to investigate the capability and accuracy of the D-Wave quantum system to solve traveling salesman problems (TSPs) that are too large to process on the DW_2000Q_6 quantum machine.  We focus on asymmetric traveling salesman problems (ATSPs) [18-20].  We study the two smallest ATSPs in the TSPLIB using D-Wave's hybrid solvers Kerberos [4] and LeapHybridSampler (LHS) [5].  We made six executions for each combination of ATSP/solver and computed Error Percent, defined by Eq. (2), for the shortest tour length for each combination, thereby measuring the quality of the D-Wave hybrid solutions.  The two ATSPs are br17 with 17 nodes and ftv33 with 34 nodes.

### A2.  Results

### A2.1.  br17 An Asymmetric Traveling Salesman Problem

The distance matrix for br17 has several peculiarities. The smallest distance is 0. There are 31 0's in the matrix. Each row and column have at least one 0. Five rows in the matrix are identical to the row following it, except for the diagonal entries. An unusually large number of successive entries are identical. There is no indication how the matrix originated or who created it. The matrix was received by Bruno Repetto.

The largest entry in the distance matrix for br17 is 74. The shortest tour length for br17 is 39 according to [1] TSPLIB, Documentation > Table 2. An optimal tour is not listed. We used the default parameters for the solvers and set the tunable parameter gamma to N x max($D_{xy}$) / 2 = 629 per [9]. Error Percent, defined by Eq. (2), for the Kerberos solver is 2.56% and for the LHS solver is 41.03%. Five of the six Kerberos executions found a tour of length 40 which is 1 unit above the optimal length of 39. This excellent result is mainly attributed to the unusual features of the distance matrix that are described in the previous paragraph.

### A2.2. ftv33 An Asymmetric Traveling Salesman Problem

The distance matrix for ftv33 has several patterns that span rows. This seems to imply that the matrix was not randomly generated. We did not find an indication how the matrix originated or who created it. The matrix was received by M. Fischetti.

The largest entry in the distance matrix is 332 and the smallest is 7. The shortest tour length for ftv33 is 1286 according to [1] TSPLIB, Documentation > Table 2. An optimal tour is not listed. We used the default parameters for the solvers and set the tunable parameter gamma to N x max($D_{xy}$) / 2 = 5644 per [9]. We adjusted the value of gamma but could not improve the results. Error Percent, defined by Eq. (2), for the Kerberos solver is 54.51% and for the LHS solver is 154.67%, both of which indicate that the distance matrix is difficult for D-Wave hybrids to solve.

### A3. Comparison with a Classical Result

Ahmed in [18] Table 5 reports an Error Percent of 0% using a different solution method for the two ATSPs that we examined. This implies that D-Wave solvers Kerberos and LHS need improvement so their solutions of the two ATSPs have better quality, i.e., are much closer to optimal. Since the D-Wave solvers contain trade secrets, we are not able to recommend improvements to them.

### A4. Discussion and Conclusions

Table A1 is a summary of the work we report for two ATSPS.

Table A1. Results for Two Asymmetric Traveling Salesman Problems

| ATSP | Optimal Solution | Best Kerberos Solution | Error Percent % | Best LeapHybrid Sampler Solution | Error Percent % |
|---|---|---|---|---|---|
| br17 | 39 | 40 | 2.6 | 55 | 41.0 |
| ftv33 | 1286 | 1987 | 54.5 | 3275 | 154.7 |

Optimal solutions are from [1] TSPLIB, Documentation > Table 2. Error Percent is calculated according to Eq. (2).

    All tour lengths for each ATSP handled by Kerberos were less than the shortest tour length for LeapHybridSampler (LHS). This means for the tested ATSPs, the solver Kerberos is superior to the solver LeapHybridSampler (LHS). This is evident from the Error Percentages in Table A1 and is consistent with a similar result in the main body of this paper for symmetric TSPs.

    As the number of nodes increased from 17 to 34 for ASTPs, Error Percent, defined by Eq. (2), increased for both solvers. This result is also consistent with a conclusion in the main body of this paper for symmetric TSPs.

    For the examples studied in this paper with two solvers, the conclusions for symmetric TSPs and asymmetric TSPs are very much alike, i.e., both types of TSPs have similar characteristics when analyzed by the two solvers.